**Title: Broken inversion symmetry in van der Waals topological ferromagnetic metal iron germanium telluride**

*Kai-Xuan Zhang,* Hwiin Ju, Hyuncheol Kim, Jingyuan Cui, Jihoon Keum, Je-Geun Park,* and Jong Seok Lee**

Dr. Kai-Xuan Zhang, Hyuncheol Kim, Jingyuan Cui, Jihoon Keum, Prof. Je-Geun Park
Department of Physics and Astronomy, Seoul National University, Seoul 08826, South Korea
E-mail: zaal@mail.ustc.edu.cn; jgpark10@snu.ac.kr

Dr. Kai-Xuan Zhang, Hyuncheol Kim, Jingyuan Cui, Jihoon Keum, Prof. Je-Geun Park
Center for Quantum Materials, Department of Physics and Astronomy, Seoul National University, Seoul 08826, South Korea

Dr. Kai-Xuan Zhang, Prof. Je-Geun Park
Institute of Applied Physics, Seoul National University, Seoul 08826, South Korea

Hwiin Ju, , Prof. Jong-Seok Lee
Department of Physics and Photon Science, Gwangju Institute of Science and Technology (GIST), Gwangju 61005, South Korea
E-mail: jsl@gist.ac.kr

**Abstract:** Inversion symmetry breaking is critical for many quantum effects and fundamental for spin-orbit torque, which is crucial for next-generation spintronics. Recently, a novel type of gigantic intrinsic spin-orbit torque has been established in the topological van-der-Waals (vdW) magnet iron germanium telluride. However, it remains a puzzle because no clear evidence exists for interlayer inversion symmetry breaking. Here, we report the definitive evidence of broken inversion symmetry in iron germanium telluride directly measured by the second harmonic generation (SHG) technique. Our data show that the crystal symmetry reduces from centrosymmetric $P6_3/mmc$ to noncentrosymmetric polar $P3m1$ space group, giving the three-fold

SHG pattern with dominant out-of-plane polarization. Additionally, the SHG response evolves from an isotropic pattern to a sharp three-fold symmetry upon increasing Fe deficiency, mainly due to the transition from random defects to ordered Fe vacancies. Such SHG response is robust against temperature, ensuring unaltered crystalline symmetries above and below the ferromagnetic transition temperature. These findings add crucial new information to our understanding of this interesting vdW metal, iron germanium telluride: band topology, intrinsic spin-orbit torque and topological vdW polar metal states.

**Main Text:** Symmetry constitutes the cornerstone of modern condensed matter physics since symmetry breaking would trigger exotic phase transitions and produce emergent quantum phenomena. For example, spontaneous symmetry breaking can ignite the nematic phase[1,2], superfluid[3,4], BCS-governed superconductor[5,6], etc.; breaking time-reversal symmetry leads to the paramagnetic-to-ferromagnetic transition[7] and the unconventional superconductivity[8]; inversion symmetry breaking results in nonreciprocal electronic transport[9,10], nonlinear Hall effect[11,12], nonlinear optical response[13-15], topological Weyl semimetals[16-18], etc. Therefore, tailoring symmetry by symmetry engineering provides a promising and powerful way to control matter's quantum characteristics and functionalities.

Currently, van der Waals (vdW) magnets[19-23] are one of the most active research fields and have especially been found to be promising platforms for the application of future spintronics[23,24]. At the center of its activities lies $Fe_3GeTe_2$ (FGT)[25], a rare vdW metallic magnet, and, most importantly, it has a topologically protected nodal line[26]. Thanks to the large Berry curvatures of its topological bands, FGT displays giant anomalous Hall current[26] and anomalous Nernst effect[27]. Another equally exciting development is that massive intrinsic spin-orbit torque[28,29] by charge current has been revealed in a single FGT without any heavy-metal layer, making it an ideal and unique model system for new conceptual spin-orbit torque and spintronics.

In general, spin-orbit torque requires inversion symmetry breaking to generate spin polarization by current, but pristine FGT hosts a hexagonal structure with the centrosymmetric space group $P6_3/mmc$. Previous work[29] explains this anomaly because the onsite intralayer coupling is several orders of magnitude stronger than the interlayer coupling, dominating the spin-orbit torque. However, a more natural explanation would be the inversion symmetry already broken, however small, in the bulk sample, leading to a very unusual form of the Ginzburg-Landau free energy. Yet, there is no crucial direct and conclusive evidence of broken inversion symmetry in the bulk FGT. Suppose inversion symmetry is broken by reducing the space group to a noncentrosymmetric one in bulk FGT; it will add an additional contribution to its spin-orbit torque besides the strong intralayer coupling scenario. Moreover, more emergent phenomena and rich physical properties would ensue in FGT with both time-reversal and inversion symmetry breaking: for example, nonreciprocal electronic transport[9,10], nonlinear Hall effect[11,12], nonlinear optical response[13-15], etc. A close relative system $Fe_{2.5}Co_{2.5}GeTe_2$[30,31] hosting the inversion symmetry breaking has already featured consequent polar metal states, skyrmion lattice and topological Hall effect. These two curiosities prompt us to examine the possible inversion symmetry breaking of FGT directly by measuring the second-order nonlinear optical response.

In this work, we investigate the symmetry of FGT by the second harmonic generation (SHG) technique. Fe-deficient FGT exhibits a sharp three-fold SHG pattern with dominant out-of-plane polarization, implying the breaking of inversion symmetry. By examining the pristine FGT's symmetries and performing the group-subgroup symmetry reduction analysis, we find the polar space group $P3m1$ to be the most likely new space group induced by Fe vacancies. Moreover, as Fe deficiency increases, SHG response gradually stabilizes a sharp three-fold symmetry from an isotropic pattern, reflecting the transition from random defects to ordered Fe vacancies. Furthermore, spin ordering across the ferromagnetic transition temperature does not affect the nonlinear optical response, indicating that the crystalline symmetries maintain the same for both ferromagnetic and paramagnetic phases. Finally, based on the reduced symmetries and new space group, we discuss its potential effects on the band topology, intrinsic spin-orbit torque and possible vdW polar metal states for FGT.

***Inversion symmetry breaking in $Fe_{2.8}GeTe_2$:*** We first check the inversion symmetry breaking of Fe-deficient $Fe_{2.8}GeTe_2$ by the SHG technique. Figure 1a illustrates the measurement geometry, where an 800 nm laser shines the sample with an incident angle of 45° toward the sample surface. The emitted 400 nm light ($2\omega$) is collected by a photomultiplier tube. We found that the light intensity detected shows a quadratic dependence on an incident laser power (Fig. S3), verifying that the signal is generated from the second-harmonic process. The sample is rotated in the plane during the measurement, which helps to map out its inherent symmetry. SHG pattern is obtained for the PP (SS) mode, where the incident and emitted light share the same polarization of 45° (0°) toward the sample surface. Such a general geometry can probe both the out-of-plane and in-plane crystalline polarization or symmetry breaking, if any. As shown in Fig. 1b, the SHG pattern of the PP mode clearly reveals three petals with considerable intensity. By contrast, the SS-mode SHG's intensity is extremely low—this SHG result is the definitive evidence of broken inversion symmetry in $Fe_{2.8}GeTe_2$, dominantly along the out-of-plane direction.

***Group-subgroup symmetry reduction analysis:*** To better understand the SHG results, we first examine the original symmetries of pristine $Fe_3GeTe_2$ and then perform group-subgroup symmetry reduction analysis on Fe-deficient $Fe_{3-x}GeTe_2$. As depicted in Fig. 2a, $Fe_3GeTe_2$ monolayer (A-layer) hosts the noncentrosymmetric point group $P\bar{6}m2$ with three symmetries: three-fold rotation symmetry $C_3$, mirror plane symmetries $m_z$ and $m_y$. However, the neighboring layer (B-layer) forms the inversion partner of the A-layer through the symmetries of $6_3$ and glide $c$ with 1/2 unit-cell upshift. Eventually, the entire $Fe_3GeTe_2$ with both A and B layers crystallizes in a hexagonal structure with centrosymmetric space group $P6_3/mmc$. The top view of the $Fe_3GeTe_2$ structure illustrates better $m_y$ and glide $c$ symmetries in Fig. 2b. The arrows in Fig. 2b indicate that the $Fe^{II}$, Ge, and Te atoms (represented in Fig. 2a along the vertical solid black lines) satisfy the symmetries of $6_3$ (rotate $2\pi/6$ then upshift 3/6 unit-cell) and glide $c$ with 1/2 unit-cell upshift.

Figure 2c plots the group-subgroup relationship, starting from the original $P6_3/mmc$ centrosymmetric space group of perfect $Fe_3GeTe_2$ with three constraints: primitive lattice $P$; three-fold structure; absence of inversion symmetry evidenced by the considerable 3-fold SHG response, which cannot be generated from electric quadrupole contribution (Supporting Note 1.1). The SHG results in Fig. 1 reveal that the crystalline polarization or inversion symmetry breaking is predominant along the out-of-plane direction with the vanishing in-plane one. It indicates that the pristine in-plane symmetry like $m_y$ is not much changed, but the out-of-plane symmetries $6_3$, glide $c$, and $m_z$ should be destroyed. Therefore $P3m1$ should be the possible new space group for $Fe_{2.8}GeTe_2$, consistent with very recent work[32] by fitting an X-ray diffraction (XRD) pattern. Coincidently, the out-of-plane symmetries $6_3$ and glide $c$ are the key components that make the B-layer the inversion partner of the A-layer but have been destroyed in Fe-deficient $Fe_{2.8}GeTe_2$. Based on this new space group $P3m1$, we simulated the SHG pattern from the electric dipole contribution of the point group $3m$, which nicely matches the experimental SHG response (see the blue curve in Fig. 1b) and thus, in turn, supports the new space group $P3m1$ again.

Please note that $Fe^{II}$ vacancies should be dominant in a Fe-deficient FGT by the following considerations on both the special atoms' arrangement in FGT and the principle of charge neutrality. First, $Fe^{II}$ is exactly sandwiched by the top and bottom Te atoms, while the $Fe^{III}$ site has much more free space. As a result, it is more difficult for Fe atoms to occupy the $Fe^{II}$ sites than the $Fe^{III}$ sites during the growth for Fe-deficient samples, leaving more $Fe^{II}$ sites unoccupied and leading to more $Fe^{II}$ vacancies. Second, the principle of charge neutrality will also require more occupancy of higher-valence $Fe^{III}$ ions than lower-valence $Fe^{II}$ ions in the Fe-deficient samples, again making $Fe^{II}$ vacancies dominant. Therefore, for Fe-deficient $Fe_{3-x}GeTe_2$, $Fe^{II}$ vacancies naturally can be the main contribution to the observed inversion symmetry breaking. The $Fe^{II}$ vacancy allows breaking the out-of-plane symmetries once the $Fe^{II}$ occupancy differs between the A and B layers in Fig. 2a, which has been proposed recently[32].

***SHG evolution on Fe-deficiency and temperature:*** Next, we explore the Fe-deficiency dependence of the SHG response. Three samples with different Fe ratios ($Fe_{2.8}GeTe_2$, $Fe_{2.9}GeTe_2$, and $Fe_{3.0}GeTe_2$) have been comparatively investigated, and whose basic transport properties of corresponding nanoflake devices are summarized in Fig. S1. The temperature-dependent longitudinal resistance $R_{xx}$ exhibits a kink at its ferromagnetic transition temperature (the Curie temperatures indicated by red dashed lines in Fig. S1a). The magnetic-field-dependent transverse resistance $R_{xy}$ hosts a rectangular hysteresis loop due to its significant ferromagnetic anomalous Hall effect below the Curie temperature (Fig. S1b). As expected, the Curie temperature and coercivity decrease with increasing Fe deficiency (Fig. S1c)[33]. Additionally, the exact coercivity value of each nanoflake device at low temperatures (10 or 20 K) reasonably agrees with the previous report[33] on $Fe_{3-x}GeTe_2$.

Figure 3a-c shows the normalized SHG results concerning the minimum intensity for $Fe_{2.8}GeTe_2$, $Fe_{2.9}GeTe_2$, and $Fe_{3.0}GeTe_2$, respectively. The SHG pattern gradually evolves from a well-resolved sharp 3-fold symmetry of $Fe_{2.8}GeTe_2$ to a blunt 3-fold symmetry of $Fe_{2.9}GeTe_2$, eventually toward a more isotropic pattern of $Fe_{3.0}GeTe_2$. Figure 3d demonstrates the relative SHG sharpness more quantitatively, defined as $\max(I_{SHG})/\min(I_{SHG})-1$, corresponding to 5.3, 1.5, 1 for $Fe_{2.8}GeTe_2$, $Fe_{2.9}GeTe_2$, and $Fe_{3.0}GeTe_2$, respectively. Based on these observations, we have made a general evolution picture for the SHG pattern and the symmetry reduction: even for $Fe_{3.0}GeTe_2$, random defects can exist in the single crystal, leading to the more isotropic SHG pattern. Upon increasing Fe deficiency in $Fe_{2.9}GeTe_2$, ordered Fe vacancy appears, mixing with the random defects, resulting in the blunt 3-fold symmetry of the SHG pattern. With further increasing Fe deficiency to $Fe_{2.8}GeTe_2$, ordered Fe vacancy dominates the system and thus stabilizes the sharp 3-fold symmetry, eventually reducing the crystalline symmetry to a new space group of $P3m1$. Such Fe-deficiency dependence of the bulk FGT's SHG response reinforces the scenario of Fe-vacancy-induced inversion-symmetry-breaking as the primary mechanism rather than a possible surface contribution, consistent with recent work by the fitting of XRD pattern on bulk FGT[32].

We also checked the temperature dependence of SHG response for $Fe_{2.8}GeTe_2$. As described in Fig. 4a, we fixed the sample angle so that S-polarized light is aligned normal to the $m_y$ mirror plane of a sample while fixing or rotating the polarization of incident light. The SHG intensity obtained in the PP mode (indicated by the star in Fig. 4b) remains almost constant across the whole temperature range, with a base temperature far below the ferromagnetic transition temperature $T_c$ of ~161 K. In addition, Figure 4b shows the SHG pattern with rotating the light polarizations at temperatures above (250 K), below (150 K), and far below (80 K) the ferromagnetic Curie temperature of ~161 K, respectively. The SHG patterns are almost intact when the temperature crosses the Curie temperature, indicating that the new space group $P3m1$ stabilizes both in the paramagnetic and ferromagnetic phases for $Fe_{2.8}GeTe_2$. This temperature independence is universal, as reproduced on another FGT sample (Fig. S2). Such robustness against temperature implies that FGT shares the same crystalline symmetries for both the paramagnetic and the ferromagnetic phases. It simplifies the following discussions on topological bands and intrinsic spin-orbit torque below the ferromagnetic transition temperature.

***Discussion:*** We have measured nonlinear SHG optical response to provide direct and conclusive evidence of inversion symmetry breaking in FGT. The new space group is analyzed to be $P3m1$, consistent with the previous report[32], reinforcing the Fe-vacancy-induced inversion-symmetry-breaking scenario. Our discovery requires a renewed understanding of FGT's properties from three perspectives: topological bands, intrinsic spin-orbit torque, and polar metal states.

FGT is a topological ferromagnetic nodal-line material with large Berry curvature and thus hosts consequent giant anomalous Hall current[26], anomalous Nernst effect[27], and gigantic intrinsic spin-orbit torque[29]. The screw symmetry $6_3$ is essential to ensure its topological bands[26] but has been destroyed by Fe vacancy with a new space group of $P3m1$. Strictly speaking, the topological bands cannot be perfectly maintained. However, we note that the symmetry of FGT is only weakly broken by Fe vacancy with two pieces of evidence: one is that $Fe^{II}$ occupancy for A-layer and B-layer only slightly changes from 1:1 to ~0.92:0.87 in a recent report[32]; another is that the anomalous Hall effect is still significant in all our samples, similar to previous work[26]. Therefore, the energy gap opened by such weak symmetry breaking should be small, probably smaller than the energy scale of spin-orbit coupling; which would render the topological bands and large Berry curvatures much unchanged.

Recent investigations[29,34] discover the gigantic intrinsic spin-orbit torque by current in a single FGT without any heavy-metal layer, which has also been confirmed by sequential works[35-39] from different research groups. But in principle, spin-orbit torque cannot be generated by current in a centrosymmetric system like FGT since inversion symmetry breaking is required for producing the current-driven spin polarization. Specifically for FGT, the A-monolayer can produce a spin-orbit torque. Still, its inversion partner B-layer can simultaneously produce the same spin-orbit torque of the opposite sign, eventually cancelling out to zero net torque. To understand this contradiction, previous work[29] regards it as hidden spin-orbit torque, similar to the hidden Rashba effect in centrosymmetric systems[40]. It is based on the fact that the onsite intralayer coupling (~1 eV) is three orders of magnitude stronger than the interlayer coupling (~1 meV) and thus regulates the spin-orbit torque of FGT. Moreover, the energy of a multilayer FGT is almost the same regardless of whether neighboring layers are ferromagnetically or antiferromagnetically ordered, indicating that the interlayer magnetic coupling is exceptionally weak, and the dynamics of an individual layer is almost unaffected by the dynamics of its neighboring layers[29]. Aside from this hidden spin-orbit torque explanation focusing on the coupling's energy scale, our work provides an additional contribution of inversion symmetry breaking in the actual FGT material.

Another noteworthy point is that the new space group $P3m1$ is polar, and $Fe_{3-x}GeTe_2$ can, in principle, exhibit an out-of-plane polarization, which would make it a fascinating, still much-undeveloped case of a potential vdW polar metal with its band topology. It would thus enrich future opportunities to exploit its vdW topological ferromagnetic polar metal states with the high-order harmonic electrical response, etc., as studied in the $Fe_{2.5}Co_{2.5}GeTe_2$ system[30,31] with no inversion symmetry. In addition, our results can be naturally extended to other vdW systems sharing similar pristine space groups like $P6_3/mmc$ or layer inversion, e.g., 2H-transition metal dichalcogenide (2H-TMDC), since the screw axis symmetry and layer inversion are sensitive to defects and fragile enough to be broken. As a consequence, such a general inversion symmetry breaking can explain the contradictions in many centrosymmetric systems of layer inversion induced by the inversion-symmetry-breaking-required phenomena such as chiral spin texture[32,41,42] and spin-orbit torque[29,34,43].

Here, we would like to explicitly address the novelty and importance of work. Firstly, FGT has intrinsic spin-orbit torque, which requires inversion symmetry breaking, but FGT's structure was previously thought to be centrosymmetric. Therefore, several possibilities have been suggested for breaking the inversion symmetry in FGT: for example, FGT heterostructures or FGT on substrates[28] may break FGT's inversion symmetry. In sharp contrast, our present work directly and conclusively demonstrates that FGT's inversion symmetry is inherently broken in the Fe-deficient and Fe-nondeficient FGT itself, using the sensitive nonlinear optical technique SHG. It can give an unambiguous answer to the issue of the "existence of FGT's inversion symmetry breaking and its origin". Such direct and conclusive experimental evidence of the broken inversion symmetry in FGT is critical to the whole van-der-Waals magnet society, especially around the FGT-like material family. Moreover, such inherent inversion symmetry breaking facilitates device applications since it does not have additional requirements from the substrates.

Secondly, the symmetry-breaking dependence on Fe deficiency and temperature has never been reported. Thirdly, once the inversion symmetry is broken, more emergent phenomena and rich physical properties would be followed in FGT: those new studies can be numerous, including nonreciprocal electronic transport, nonlinear Hall effect, and nonlinear optical response, to name only a few. Armed with this new correct information about the inversion symmetry breaking of FGT, we can also systematically discuss its effects on band topology, chiral spin texture, possible polar metallic states, and intrinsic SOT. Such broad implications and related intriguing properties induced by inversion symmetry breaking haven't been well aware of before in these systems.

Finally, we would like to expand our ideas further that such inversion symmetry breaking can be a more general behavior for a system with layer inversion or screw axis symmetry since these symmetries are very sensitive to defects and fragile enough to be broken. Note that numerous vdW materials like the famous 2H-TMDC also share the same space group of $P6_3/mmc$, meaning that our work has much broader implications beyond FGT itself. Our prototypical work can explain the contradictions in other centrosymmetric systems of layer inversion induced by the inversion-symmetry-breaking-required phenomena. To summarize, the novelty and importance of our work cover many exciting facets of this vdW ferromagnet FGT: the existence and inherent origin of inversion symmetry breaking, inversion symmetry breaking's evolution, the comprehensive discussions on its effects to many quantum properties and also the outlooks to many similar systems.

In summary, using the nonlinear optical response SHG as a sensitive probe, we provide conclusive evidence for the broken inversion symmetry in an important vdW metallic magnet FGT. Upon increasing Fe deficiency, the symmetry reduces from the original centrosymmetric $P6_3/mmc$ space group to the noncentrosymmetric polar space group $P3m1$, stabilizing a sharp three-fold SHG pattern with dominant out-of-plane polarization. The symmetry and corresponding SHG response remain unchanged across the ferromagnetic transition from high to low temperatures. Most importantly, it provides more insights and opportunities for vdW FGT and other layered materials of similar space groups in the fields of topological materials and spintronics via symmetry engineering.

**Experimental Section**

*Growth of FGT single crystals:* FGT single crystals were grown by the conventional chemical vapour transport method with iodine as the transport agent, following our previous works[44,45]. Pure element powders of Fe, Ge, and Te were mixed in varying ratios for different samples and sealed in a quartz tube under vacuum. Afterwards, we placed the tube in a two-zone furnace for growing single crystals, with a source and sink temperature of 750 and 650 ˚C, respectively, for seven days. Before using those crystals for further experiments, we adopted energy dispersive X-ray analysis to confirm the chemical ratio of Fe atoms in the crystals: $Fe_{2.8}GeTe_2$, $Fe_{2.9}GeTe_2$, and $Fe_{3.0}GeTe_2$, for the SHG measurements. We also fabricated corresponding nanoflake devices to check their basic transport properties.

*Electrical transport measurements:* We performed transport measurements using a resistivity probe operated inside a cryostat down to 2.5 K. The resistance was measured by using a standard lock-in technique with Stanford SR830. Gold wires were wire-bonded to connect the electronic chip to the sample's Au/Ti electrodes. An antistatic wrist strap was used during the operation to prevent the possible damage of electrostatic discharges or shocks to the sample.

*Second harmonic generation measurements:* We used 800 nm wavelength Ti-Sapphire laser pulses at an 80 MHz repetition rate. The laser beam was focused with an objective lens and incident to the sample with an incidence angle of 45°. Second harmonic light with 400 nm wavelength was generated from the sample, which was then collimated with another objective lens and collected with a photomultiplier tube (Hamamatsu). Laser intensity was modulated with a chopper to utilize lock-in detection. The polarization of both fundamental and second harmonic light was determined as p-/s- polarization with a polarizer and half-wave plate. SHG rotation patterns were obtained by rotating the sample or rotating the polarization of the incident and second harmonic light. Low-temperature measurement was performed with a cryostat (MicrostatHe, Oxford). The samples were cleaved before the measurement and kept in inert gas or vacuum. The position-dependent variations of the SHG response are negligible (Fig. S5), revealing a homogeneous macroscopic structural symmetry.

**Supporting Information**
Supporting Information is available from the Wiley Online Library or the author.
*Note added.* During the review process of our work, we came across a related work on arXiv[46].

**Acknowledgments**
K.Z. and H.J. contributed equally to this work. We thank Sang-Wook Cheong, Hyun-Woo Lee, Mijin Lim, Pyeongjae Park, and Suhan Son for their support and helpful discussions. The work at CQM and SNU was supported by the Samsung Science & Technology Foundation (Grant No. SSTF-BA2101-05) and the Leading Researcher Program of the National Research Foundation of Korea (Grant No. 2020R1A3B2079375). The work at GIST was supported by the National Research Foundation of Korea (NRF) grant funded by the Korean government (MSIT) (No. 2022R1A2C2007847). In addition, the Samsung Advanced Institute of Technology also supported this work at SNU.

Received: ((will be filled in by the editorial staff))
Revised: ((will be filled in by the editorial staff))
Published online: ((will be filled in by the editorial staff))

**Figures**

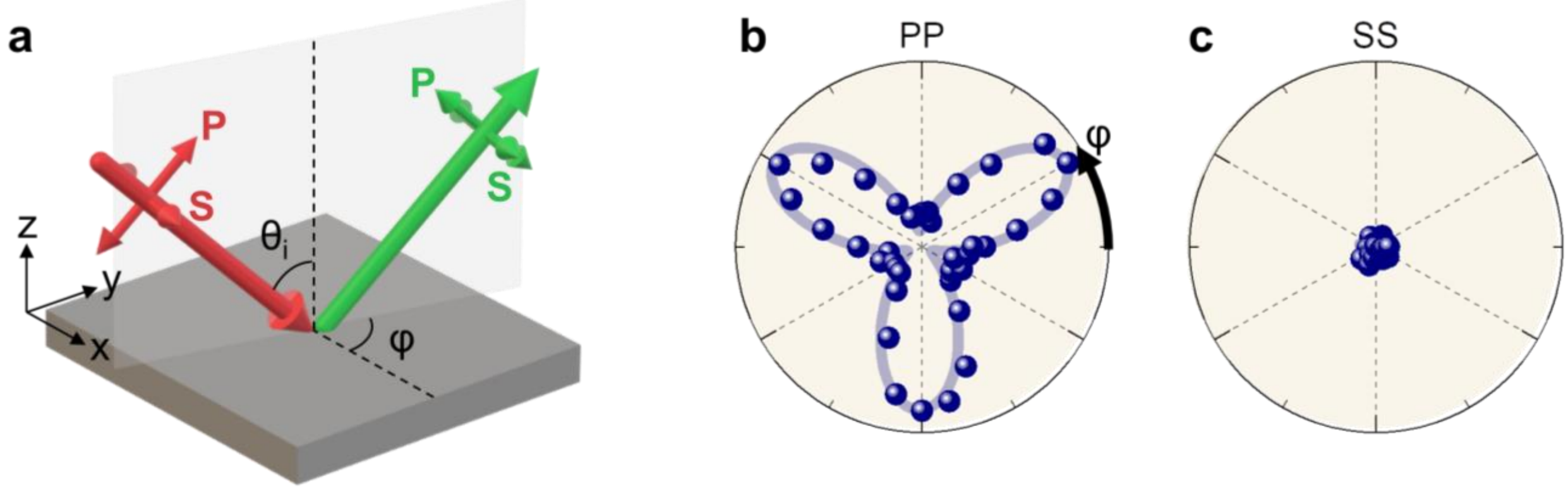

**Figure 1.** Inversion symmetry breaking in $Fe_{2.8}GeTe_2$ by SHG measurement. a) Schematic of SHG measurement. The incident (red arrow) and emitted (green arrow) light move at 45$^o$ toward the sample surface, and the sample is rotated during the measurements. The measurements adopt the PP (SS) mode, where the light polarizations are 45$^o$ (0$^o$) to the sample surface. The double-headed arrows indicate the light polarization direction. b) SHG response in the PP mode. It shows a well-resolved three-fold pattern with considerable intensity, evidencing the existence of inversion symmetry breaking. The blue curve represents the simulated SHG pattern regarding the new *3m* point group. c) SHG response in the SS mode, with extremely low intensity. It indicates the in-plane polarization is nearly zero while the out-of-plane polarization predominates in (b).

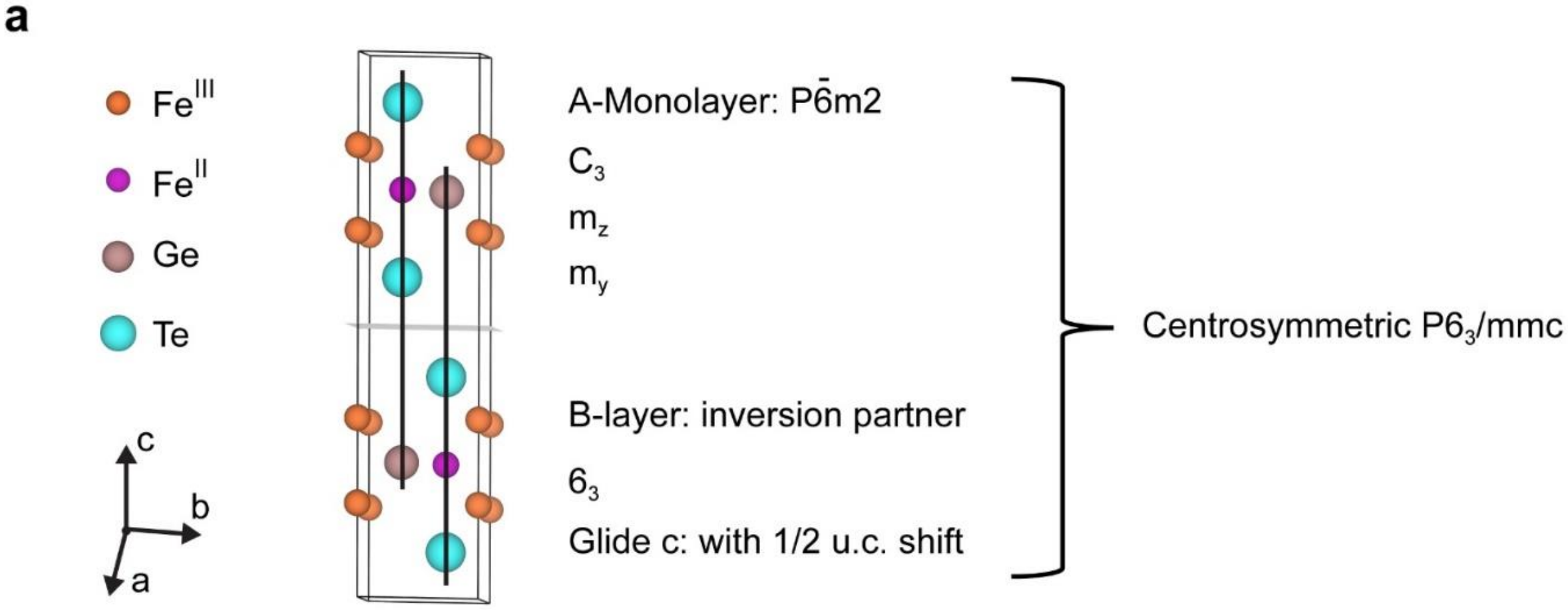

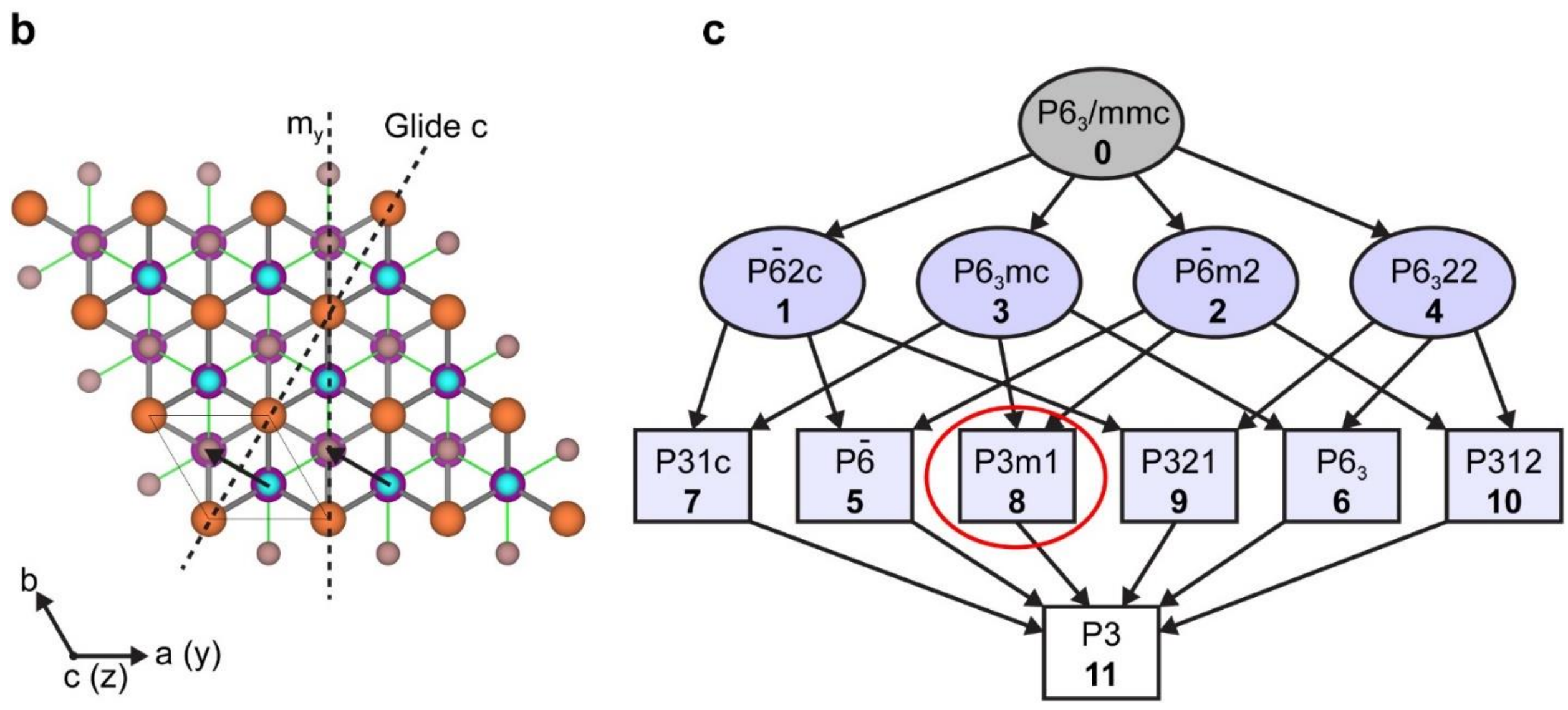

**Figure 2.** Crystalline symmetries of perfect FGT and the group-subgroup symmetry reduction analysis. a) Crystallography of a perfect FGT unit cell. The yellow, pink, brown and blue balls represent the $Fe^{III}$, $Fe^{II}$, Ge, and Te atoms, respectively. The cleavable (grey) plane cuts the FGT's unit cell in half with the top A-monolayer and bottom B-layer. A-layer belongs to the noncentrosymmetric point group $P\bar{6}m2$ with three symmetries: three-fold rotation symmetry $C_3$, mirror plane symmetries $m_z$ and $m_y$. B-layer forms the inversion partner of A-layer by the symmetries of $6_3$ and glide *c* with 1/2 unit-cell upshift. Eventually, the entire $Fe_3GeTe_2$ with both A and B layers crystallizes in a hexagonal structure with centrosymmetric space group $P6_3/mmc$. Please note the $Fe^{II}$, Te, and Ge atoms are located vertically on a line, and $F^{II}$ sites are sandwiched by the top and bottom Te atoms. b) Top view of the FGT's unit cell. The dashed lines mark the position of the $m_y$ mirror plane and the $c$ glide plane. The arrows indicate that the $Fe^{II}$, Ge, and Te atoms along the vertical line in (a) satisfy the symmetries of $6_3$ (rotate $2\pi/6$ then upshift 3/6 unit-cell) and glide $c$ with 1/2 unit-cell upshift. c) Group-subgroup relationship. It starts from the original $P6_3/mmc$ centrosymmetric space group of perfect FGT and highlights the new $P3m1$ space group for Fe-deficient FGT (red circle).

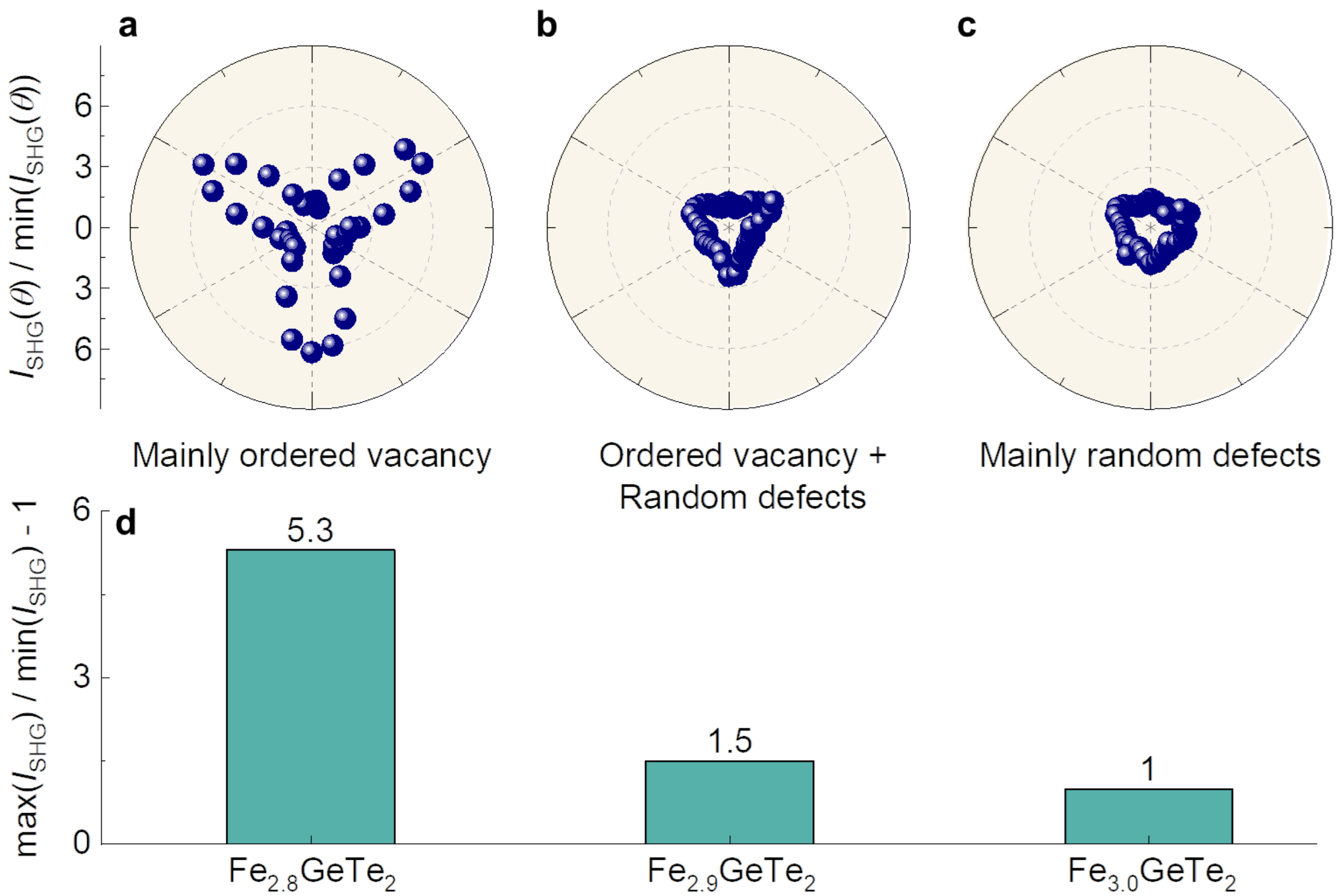

**Figure 3.** SHG evolution on Fe-deficiency. a-c) Normalized SHG response by minimum intensity for $Fe_{2.8}GeTe_2$, $Fe_{2.9}GeTe_2$, and $Fe_{3.0}GeTe_2$, respectively. d) Relative SHG sharpness, defined as max($I_{SHG}$)/min($I_{SHG}$)-1, which corresponds to 5.3, 1.5, 1 for $Fe_{2.8}GeTe_2$, $Fe_{2.9}GeTe_2$, and $Fe_{3.0}GeTe_2$, respectively. Upon increasing Fe-deficiency, SHG response evolves from an isotropic pattern to a sharp three-fold petal, reflecting the contribution's transition from random defects to ordered Fe vacancies.

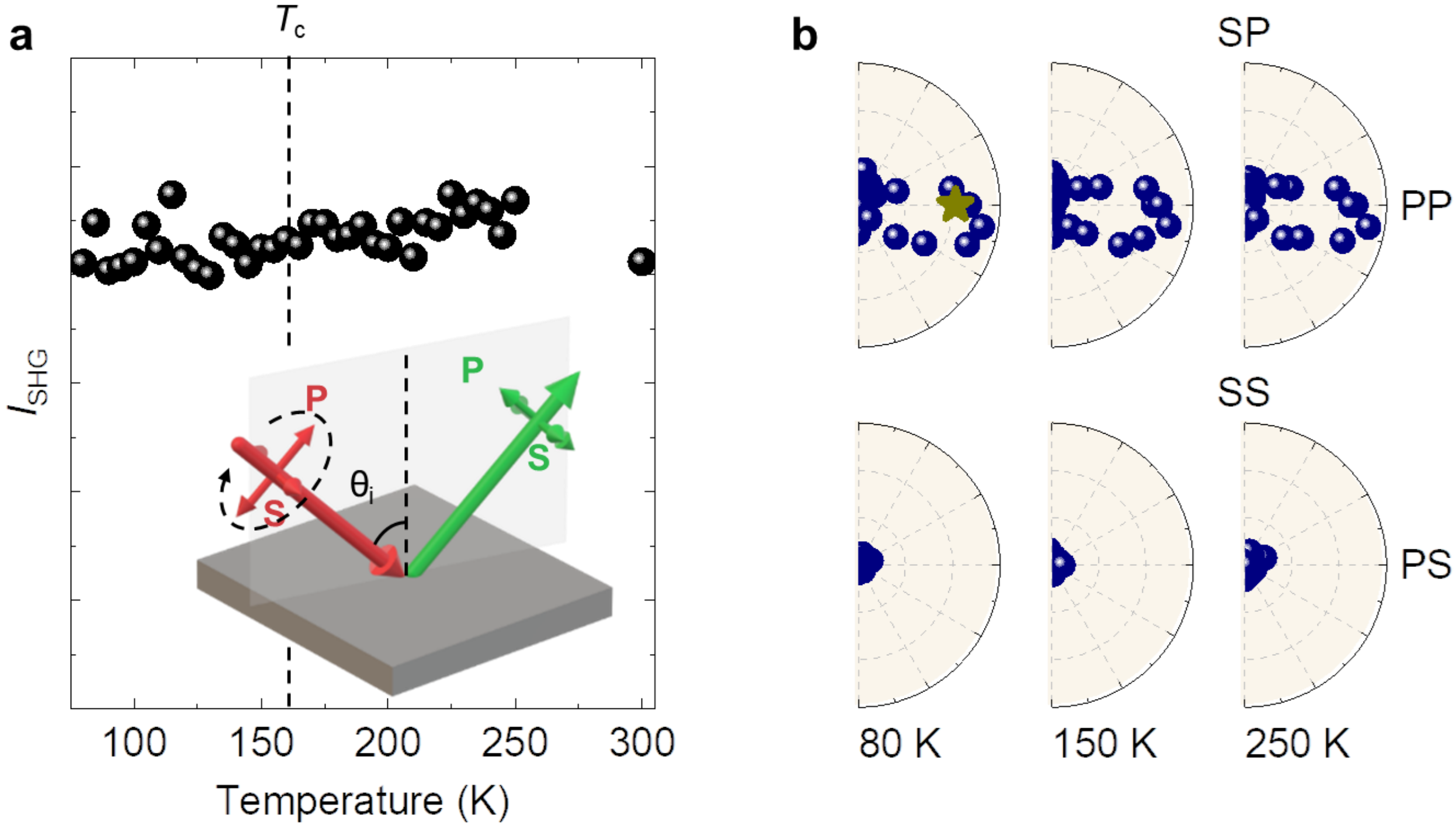

**Figure 4.** Temperature-independence of the SHG. a) SHG intensity in the PP mode (indicated by the star in (b)) as a function of temperature. It remains nearly constant across the whole temperature range, with a base temperature far below the ferromagnetic transition temperature $T_c$ of ~161 K. The inset illustrates the schematic of varying-temperature SHG measurement. The sample is anchored, and the incident light polarization is fixed to P-polarization or rotated. The emitted light is fixed to P-polarization or S-polarization. The dashed curved arrow indicates the light polarization rotation direction. b) The SHG result of rotating incident light polarization while fixing the other, for 80 K (far below Curie temperature), 150 K (below Curie temperature) and 250 K (far above Curie temperature), respectively. The SHG shows no temperature dependence. The robustness of SHG against temperature indicates that the paramagnetic and ferromagnetic phases share the same crystalline symmetries.

**The table of contents entry**

Iron germanium telluride (FGT) is a very interesting van-der-Waals ferromagnet. Despite its recent interest, a fundamental question remains unresolved about a possible broken inversion symmetry. We report the definitive evidence for such symmetry breaking in FGT using the second harmonic generation (SHG). A comprehensive investigation of Fe-deficiency and temperature-dependent SHG evolution concludes that Fe vacancies break the inversion symmetry and reduce the centrosymmetric $P6_3/mmc$ to the noncentrosymmetric polar $P3m1$. Our findings add crucial new information to understanding FGT: the band topology, intrinsic spin-orbit torque, skyrmion, and possible topological van-der-Waals polar metal states.

Kai-Xuan Zhang,* Hwiin Ju, Hyuncheol Kim, Jingyuan Cui, Jihoon Keum, Je-Geun Park,* and Jong Seok Lee*

Title: Broken inversion symmetry in van der Waals topological ferromagnetic metal iron germanium telluride

ToC figure

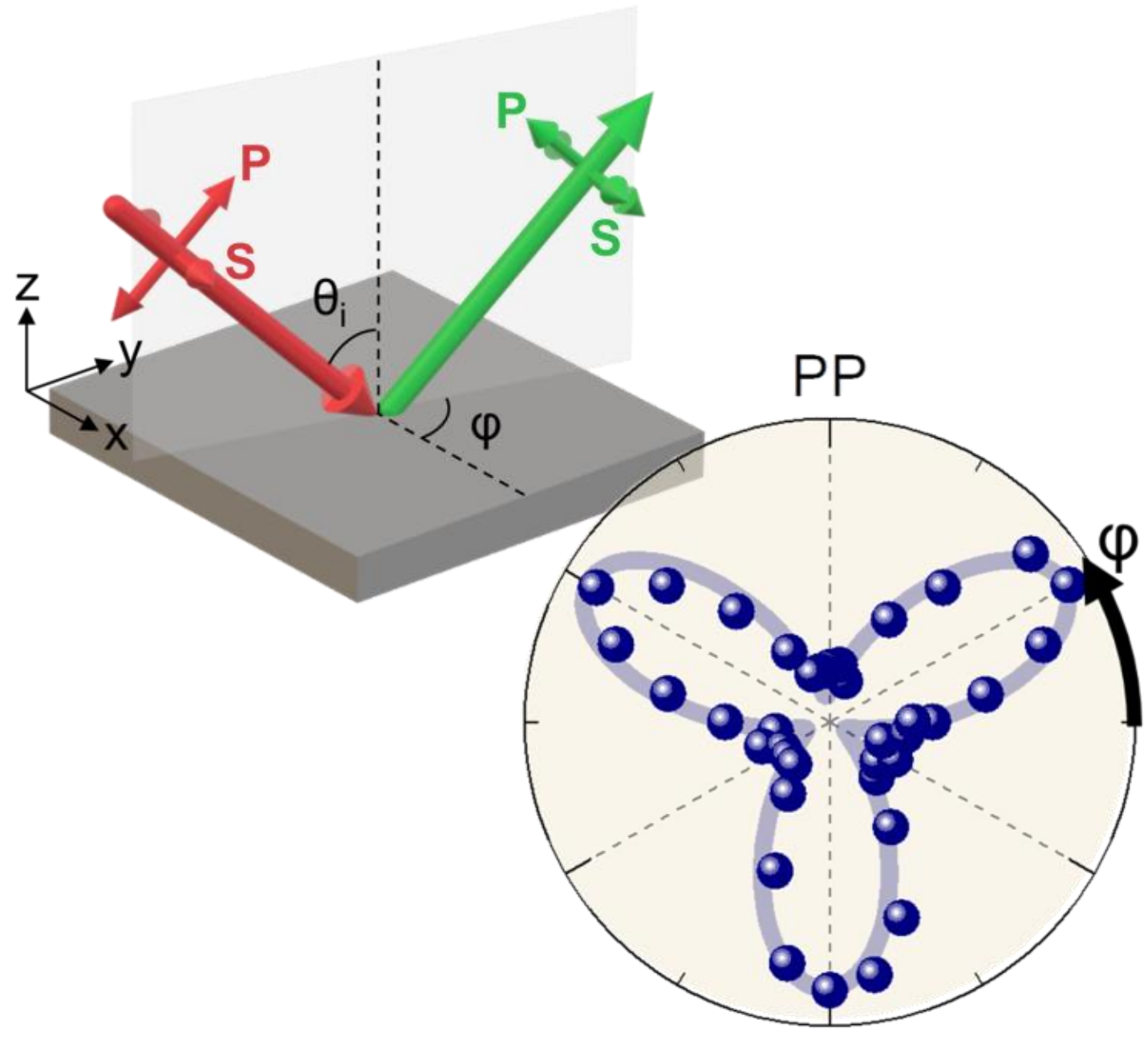

# Supporting Information

**Broken inversion symmetry in van der Waals topological ferromagnetic metal iron germanium telluride**

*Kai-Xuan Zhang,* Hwiin Ju, Hyuncheol Kim, Jingyuan Cui, Jihoon Keum, Je-Geun Park,* and Jong Seok Lee**

K.Z. and H.J. contributed equally to this work.

## Supporting Notes

### 1. **Contribution analysis for the SHG response**

#### 1.1. Electric quadrupole (EQ) contribution of the point group 6/*mmm*

Pristine $Fe_3GeTe_2$ (FGT) hosts 6/*mmm* point group structure. Since 6/*mmm* is inversion symmetric, we can neglect the electric dipole (ED) contribution. However, a possibility exists that second harmonic light is generated from electric quadrupole (EQ) contribution. The EQ-induced second-harmonic polarization $P_i^{EQ}(2\omega)$ can be expressed as

$$P_i^{EQ}(2\omega) = \chi_{ijkl} E_j(\omega) \partial_k E_l(\omega),$$

where $\chi_{ijkl}$ is third-order EQ susceptibility and $E_j(\omega)$ is the electric field of fundamental light. Subscript $i$, $j$, $k$ and $l$ indicate the Cartesian coordinates $x$, $y$, and $z$, respectively. The non-zero susceptibility components of the 6/*mmm* point group denoted by its Cartesian indices are given as

$$zzzz,$$

$$xxxx = yyyy = xxyy + xyyx + xyxy,$$

$$xxyy = yyxx = xyyx = yxxy,\ xyxy = yxyx,$$

$$yyzz = xxzz = yzzy = xzzx,$$

$$zzyy = zzxx = zyyz = zxxz,$$

$$yzyz = xzxz,$$

$$zyzy = zxzx.$$

In a 45° incidence geometry, SHG intensity generated from 6/*mmm* point group with respect to the sample angle $\phi$ can be simulated as

$$I^{2\omega}_{P_{in},P_{out}}(\phi) = \left(\frac{1}{4}\left(2\chi_{xxyy} + \chi_{xyxy} - \chi_{yyzz} + \chi_{yzyz} + \chi_{zyzy} - 2\chi_{zzyy} + \chi_{zzzz}\right)\right)^2$$

$$I^{2\omega}_{P_{in},S_{out}}(\phi) = 0$$

$$I^{2\omega}_{S_{in},P_{out}}(\phi) = \left(\frac{1}{2}\left(\chi_{xyxy} + \chi_{zyzy}\right)\right)^2$$

$$I^{2\omega}_{S_{in},S_{out}}(\phi) = 0$$

where the subscript indicates the input and output polarization of light. $I^{2\omega}_{P_{in},P_{out}}(\phi)$ and $I^{2\omega}_{S_{in},P_{out}}(\phi)$ give constant values independently of $\phi$, and $I^{2\omega}_{P_{in},S_{out}}(\phi)$ and $I^{2\omega}_{S_{in},S_{out}}(\phi)$ should be 0. Therefore, we can conclude that the EQ contribution cannot explain the 3-fold SHG results of FGT.

1.2. Electric dipole (ED) contribution of the point group 3*m*

The inversion symmetry is broken in crystals with point group 3*m,* and the second harmonic light can be generated mainly from its electric dipole (ED) contribution. The ED-induced second-harmonic polarization $P_i^{ED}(2\omega)$ can be expressed as

$$P_i^{ED}(2\omega) = \chi_{ijk}E_j(\omega)E_k(\omega),$$

where $\chi_{ijk}$ is a second-order susceptibility tensor. The non-zero susceptibility components of 3*m* point group denoted by its Cartesian indices are given as

*xzx* = *yzy*= *xxz* = *yyz, zxx* = *zyy, zzz,*

*yyy* = -*yxx* = -*xxy* = -*xyx*

In a 45° incidence geometry, SHG intensity generated from the 3*m* point group with respect to the sample angle $\phi$ can be simulated as

$$I^{2\omega}_{P_{in},P_{out}}(\phi) = \left(\frac{\sqrt{2}}{4}\left(2\chi_{xzx} - \chi_{zxx} - \chi_{zzz} + \chi_{yyy}\sin(3\phi)\right)\right)^2$$

$$I^{2\omega}_{P_{in},S_{out}}(\phi) = \left(\frac{1}{2}\chi_{yyy}\cos(3\phi)\right)^2$$

$$I^{2\omega}_{S_{in},P_{out}}(\phi) = \left(\frac{\sqrt{2}}{2}\left(\chi_{zxx} + \chi_{yyy}\sin(3\phi)\right)\right)^2$$

$$I^{2\omega}_{S_{in},S_{out}}(\phi) = \left(\chi_{yyy}\cos(3\phi)\right)^2$$

## Supporting Figures

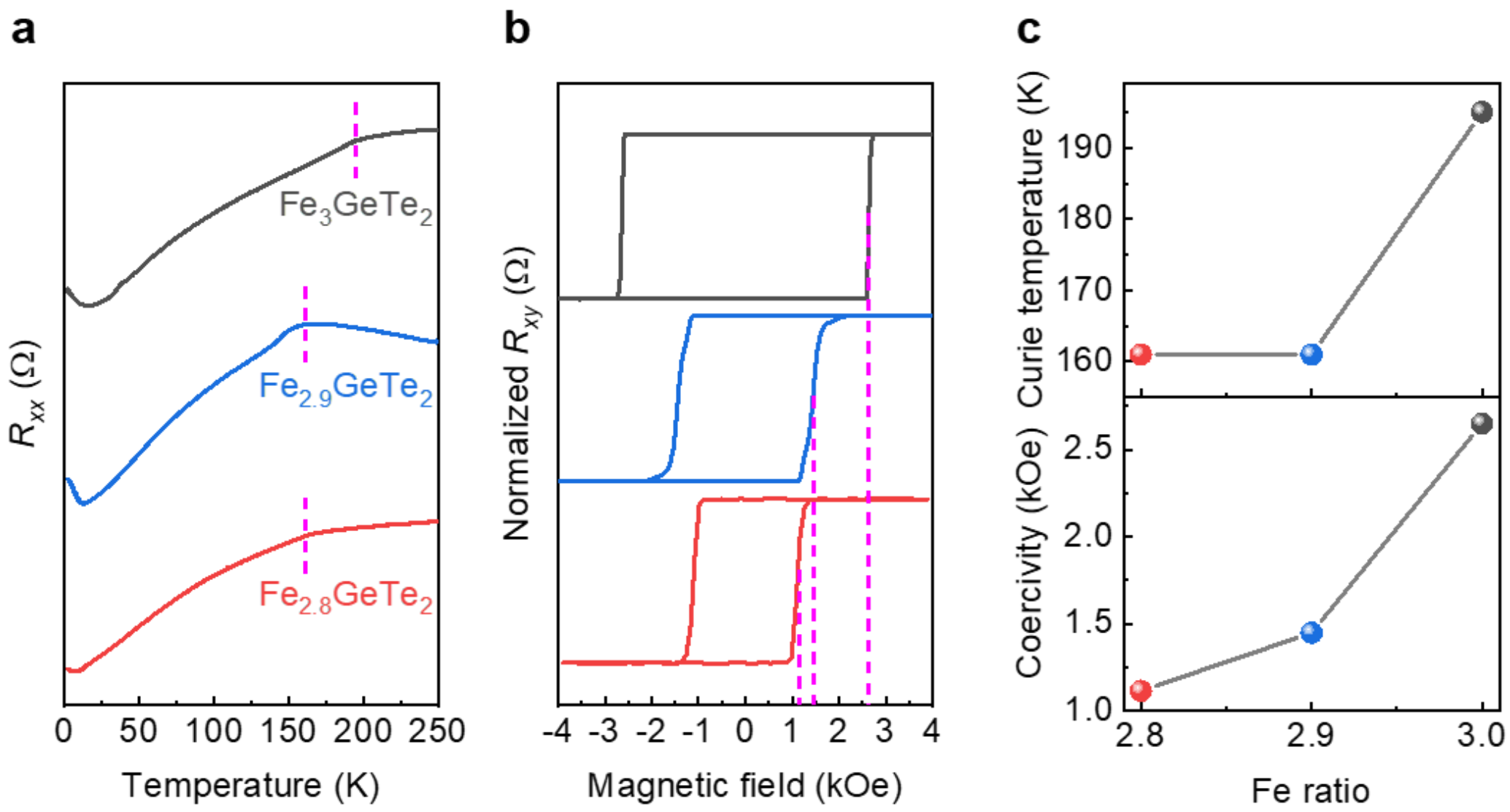

**Figure S1.** Basic transport properties of FGT devices. a) Longitudinal resistance $R_{xx}$ as a function of temperature for $Fe_3GeTe_2$ (black), $Fe_{2.9}GeTe_2$ (blue), and $Fe_{2.8}GeTe_2$ (red) nanoflakes, respectively. The pink dashed lines represent the resistance kink at its Curie ferromagnetic transition temperatures. b) Normalized transverse resistance $R_{xy}$ as a function of magnetic field for $Fe_3GeTe_2$ (black), $Fe_{2.9}GeTe_2$ (blue), and $Fe_{2.8}GeTe_2$ (red), respectively. The pink dashed lines indicate the coercive field, i.e., the coercivity of the ferromagnetic hysteresis loops due to the anomalous Hall effect. c) Curie temperature and coercivity gradually reduce on decreasing Fe ratio. The coercivity values are reasonably consistent with previous reports on $Fe_{3-x}GeTe_2$[1].

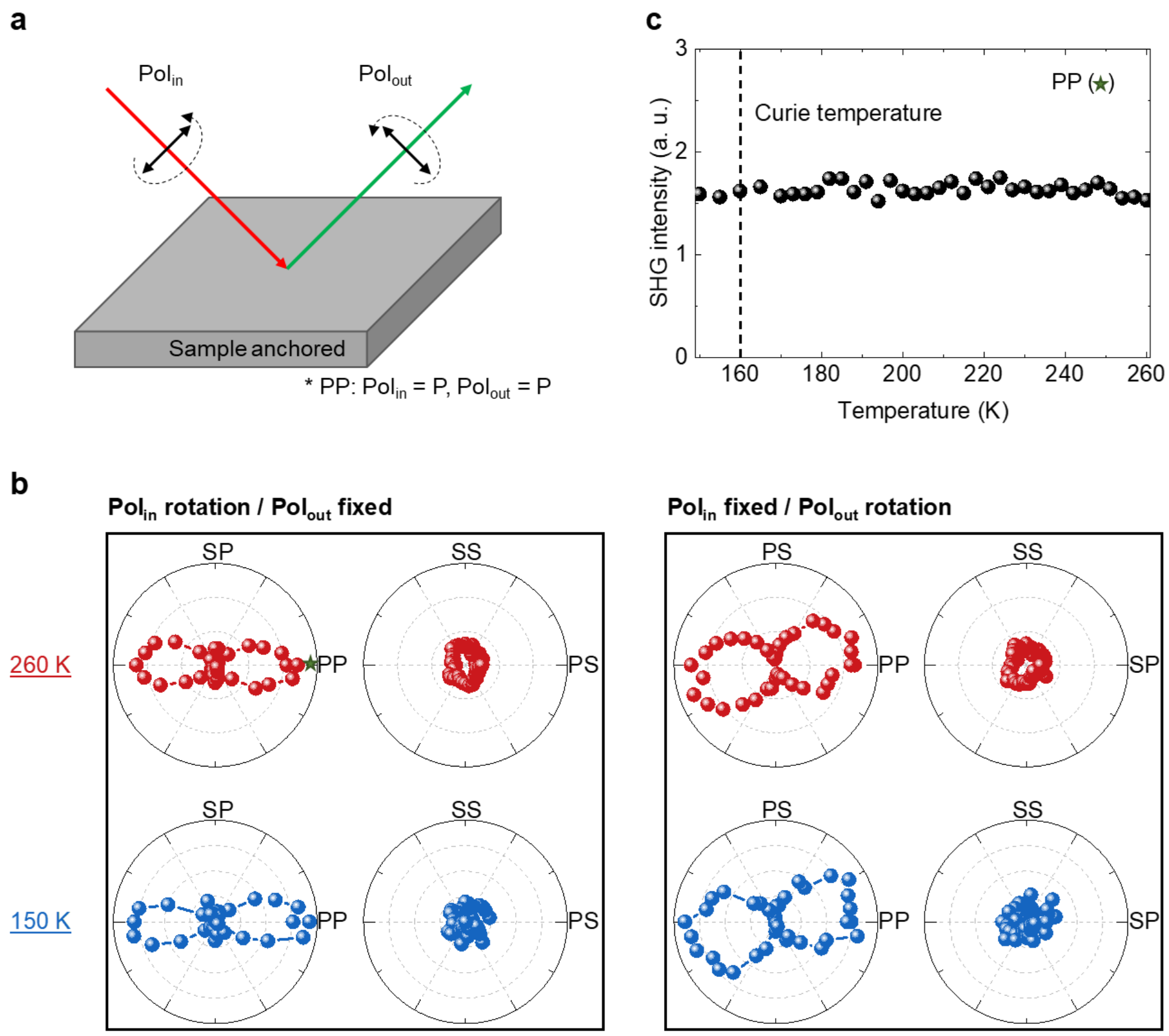

**Figure S2.** Temperature-independence of the SHG on another FGT sample. a) Schematic of varying-temperature SHG measurement. The sample is anchored, and the light polarization is rotated for the incident and emitted light. The solid double-headed arrows represent the light polarization status of the PP mode. The dashed curved arrows indicate the light polarization rotation direction. b) The SHG result of rotating incident or emitted light polarization while fixing the other for 260 K (above Curie temperature) and 150 K (below Curie temperature), respectively. The SHG shows no temperature dependence. c) SHG intensity in the PP mode (indicated by the star in (b)) as a function of temperature. It remains nearly constant across the ferromagnetic transition temperature. The robustness of SHG against temperature indicates that the paramagnetic and ferromagnetic phases share the same crystalline symmetries.

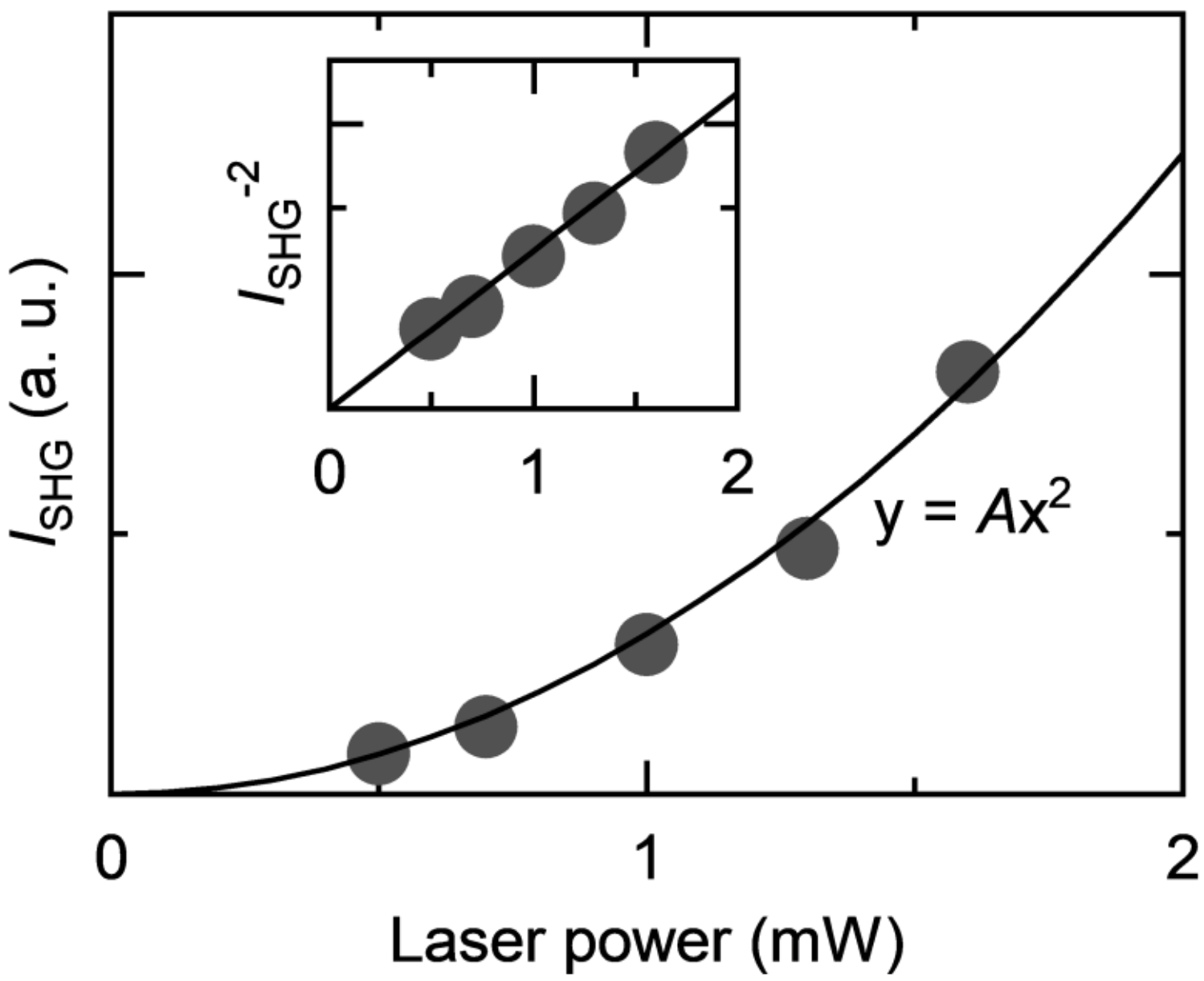

**Figure S3.** Intensity variations of second harmonic light with laser power. It shows a clear quadratic dependence, indicating that our observed signal is generated from the second harmonic process.

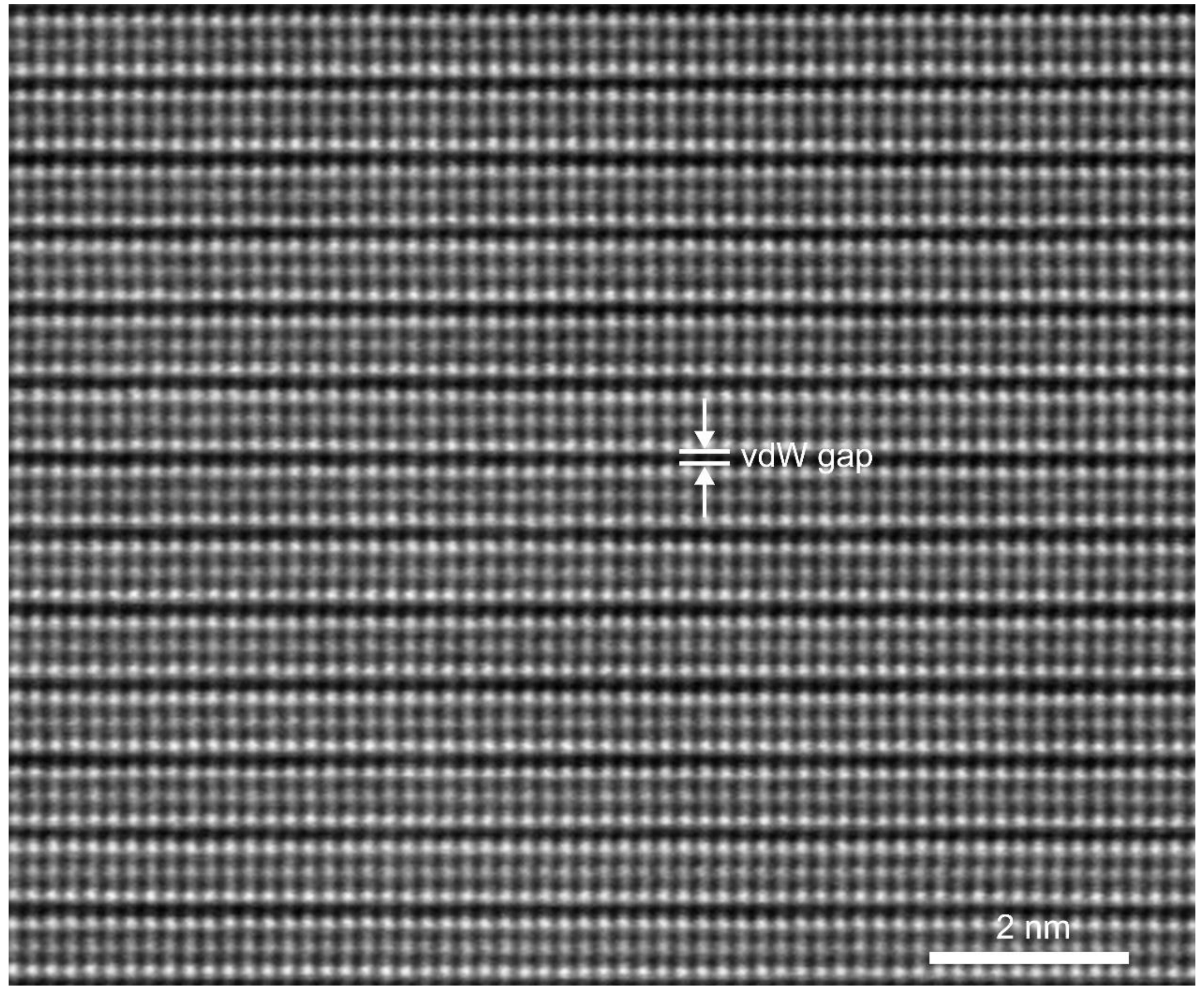

**Figure S4.** Cross-section transmission electron microscopy image of the FGT crystal. No noticeable intercalated Fe atoms are observed across the vdW gap (Figure S4 represents the full-size photo of the local zoom-in image in Fig. 1c of our previous work[2], but in a much larger scale). In principle, the statistically different Fe occupancy between the A and B layer and the possible intercalated Fe atoms within the vdW gap[3] can coexist in FGT and affect the inversion symmetry. However, we experimentally didn't observe noticeable intercalated Fe atoms within the vdW gap. Therefore, the statistically different Fe occupancy between the A and B layers can be the primary contribution to the inversion symmetry breaking, rather than the intercalated Fe atoms within the vdW gap, at least for our FGT samples.

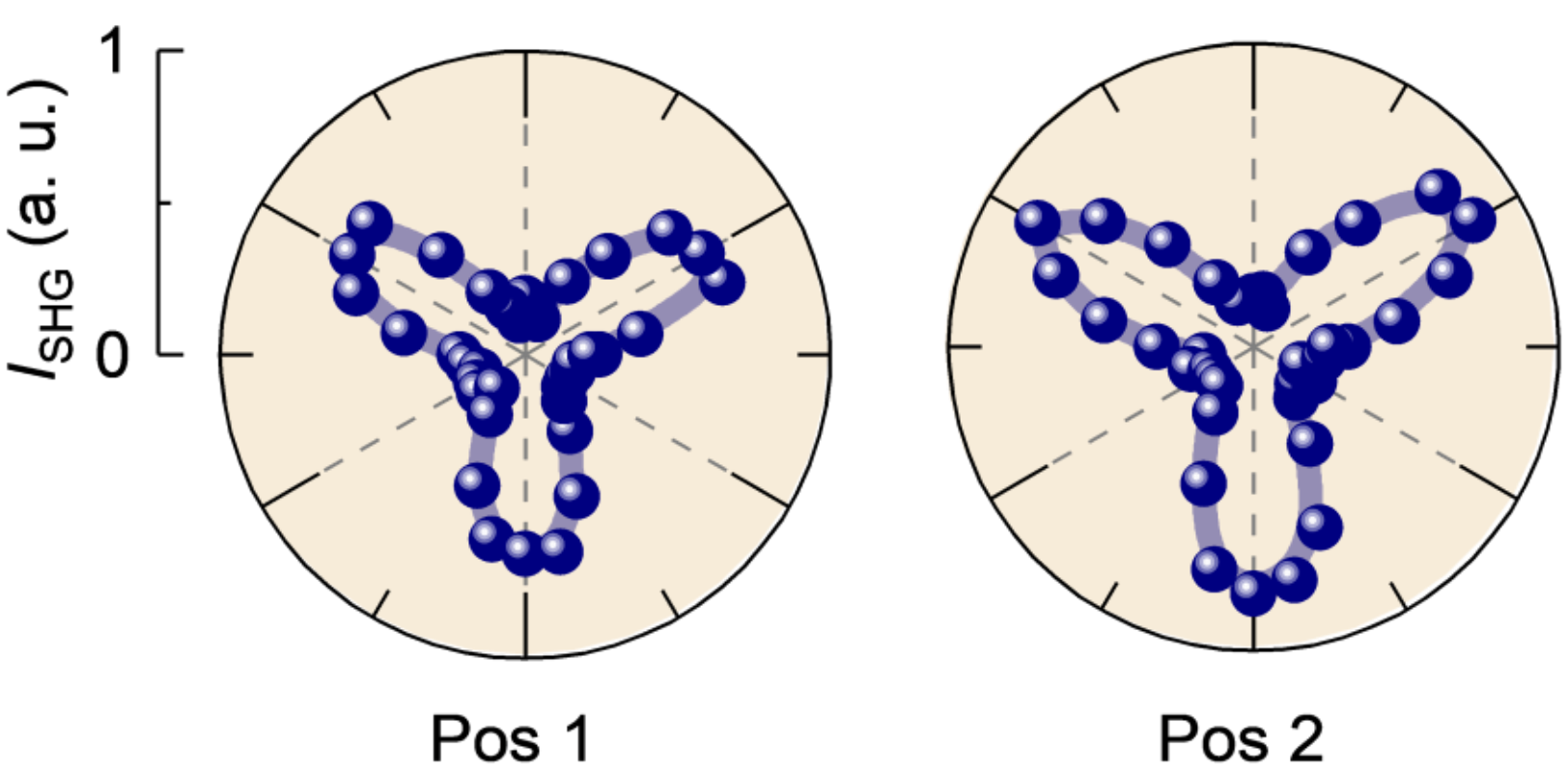

**Figure S5.** Significant SHG patterns with $P_{in}$-$P_{out}$ configuration obtained at different positions far apart. The SHG responses are more or less the same in amplitude and anisotropy patterns.

## Supporting References